\documentclass[iop,apj,numberedappendix]{emulateapj}
\usepackage{natbib}
\usepackage{graphicx}
\usepackage{amsmath,amsfonts}
\usepackage{multirow}
\usepackage{booktabs}
\usepackage{sidecap}
\usepackage{hyperref}

\slugcomment{Accepted for publication in ApJ, October 9, 2015}

\setcitestyle{notesep={; }}

\def\sgra{Sgr~A$^{\ast}$}

\def\lsim{\mathrel{\raise.3ex\hbox{$<$\kern-.75em\lower1ex\hbox{$\sim$}}}}
\def\gsim{\mathrel{\raise.3ex\hbox{$>$\kern-.75em\lower1ex\hbox{$\sim$}}}}
\def\gtwid{\mathrel{\raise.3ex\hbox{$>$\kern-.75em\lower1ex\hbox{$\sim$}}}}
\def\proptwid{\mathrel{\raise.3ex\hbox{$\propto$\kern-.75em\lower1ex\hbox{$\sim$}}}}

\begin{document}
\title{Measuring the Direction and Angular Velocity of a Black Hole Accretion Disk via Lagged Interferometric Covariance}
\shorttitle{}

\author{Michael D.\ Johnson\altaffilmark{1}, 
Abraham Loeb\altaffilmark{1}, 
Hotaka Shiokawa\altaffilmark{1,2}, 
Andrew A.~Chael\altaffilmark{1}, 
Sheperd S.\ Doeleman\altaffilmark{1,3}} 
\shortauthors{Johnson et al.}
\altaffiltext{1}{Harvard-Smithsonian Center for Astrophysics, 60 Garden Street, Cambridge, MA 02138, USA}
\altaffiltext{2}{Department of Physics \& Astronomy, The Johns Hopkins University 3400 N.\ Charles Street Baltimore, MD 21218}
\altaffiltext{3}{Massachusetts Institute of Technology, Haystack Observatory, Route 40, Westford, MA 01886, USA}
\email{mjohnson@cfa.harvard.edu}

\begin{abstract}
We show that interferometry can be applied to study irregular, rapidly rotating structures, as are expected in the turbulent accretion flow near a black hole. Specifically, we analyze the lagged covariance between interferometric baselines of similar lengths but slightly different orientations. 
For a flow viewed close to face-on, we demonstrate that the peak in the lagged covariance indicates the direction and angular velocity of the emission pattern from the flow. Even for moderately inclined flows, the covariance robustly estimates the flow direction, although the estimated angular velocity can be significantly biased.  
Importantly, measuring the direction of the flow as clockwise or counterclockwise on the sky breaks a degeneracy in accretion disk inclinations when analyzing time-averaged images alone.  
We explore the potential efficacy of our technique using three-dimensional, general relativistic magnetohydrodynamic (GRMHD) simulations, and we highlight several baseline pairs for the Event Horizon Telescope (EHT) that are well-suited to this application. 
These results indicate that the EHT may be capable of estimating the direction and angular velocity of the emitting material near \sgra, and they suggest that a rotating flow may even be utilized to \emph{improve} imaging capabilities. 
\end{abstract}


\section{Introduction}

The innermost accretion flows around black holes are the subject of intense numerical study despite a dearth of observational constraints. Even the Galactic Center supermassive black hole, Sagittarius~A$^{\ast}$ (\sgra), has yielded only limited conclusions about its accretion environment and dynamics \citep{Genzel_2010,Yuan_2014}, with x-ray observations constraining the accretion boundary conditions on scales comparable to the Bondi radius, at roughly $10^5$ times the gravitational radius $r_{\rm G} \equiv G M/c^2$ of \sgra\ \citep{Baganoff_2003}, and with radio observations providing rough estimates of the stratified size of the emission region ($20-200$ times  $r_{\rm G}$) at wavelengths from $3-13$ mm \citep{Lo_1998,Shen_2005,Bower_2006,Lu_2011,Bower_2014,Gwinn_2014}. 
Moreover, because the radio measurements are strongly affected by interstellar scattering, it was measurements of polarization and Faraday rotation that conclusively constrained the flow properties on these smaller scales \citep{Aitken_2000,Quataert_Gruzinov_2000,Bower_2003,Marrone_2007}.

This limited observational perspective will change abruptly with the completion of the Event Horizon Telescope (EHT), a project to develop a global 1.3-mm and 0.87-mm very-long-baseline interferometry (VLBI) network \citep{Doeleman_2009}. Ultimately, this network will provide a nominal angular resolution of tens of microarcseconds, sufficient to resolve the event horizons of the nearest supermassive black holes, including the Galactic Center black hole, \sgra. The EHT can achieve detections on \sgra\ with an integration time of minutes or less -- significantly shorter than the period of the innermost stable circular orbit of \sgra\ -- rendering it sensitive to both steady features and variability in the emission from \sgra.

\begin{figure}[t]
\centering
\includegraphics*[width=0.45\textwidth]{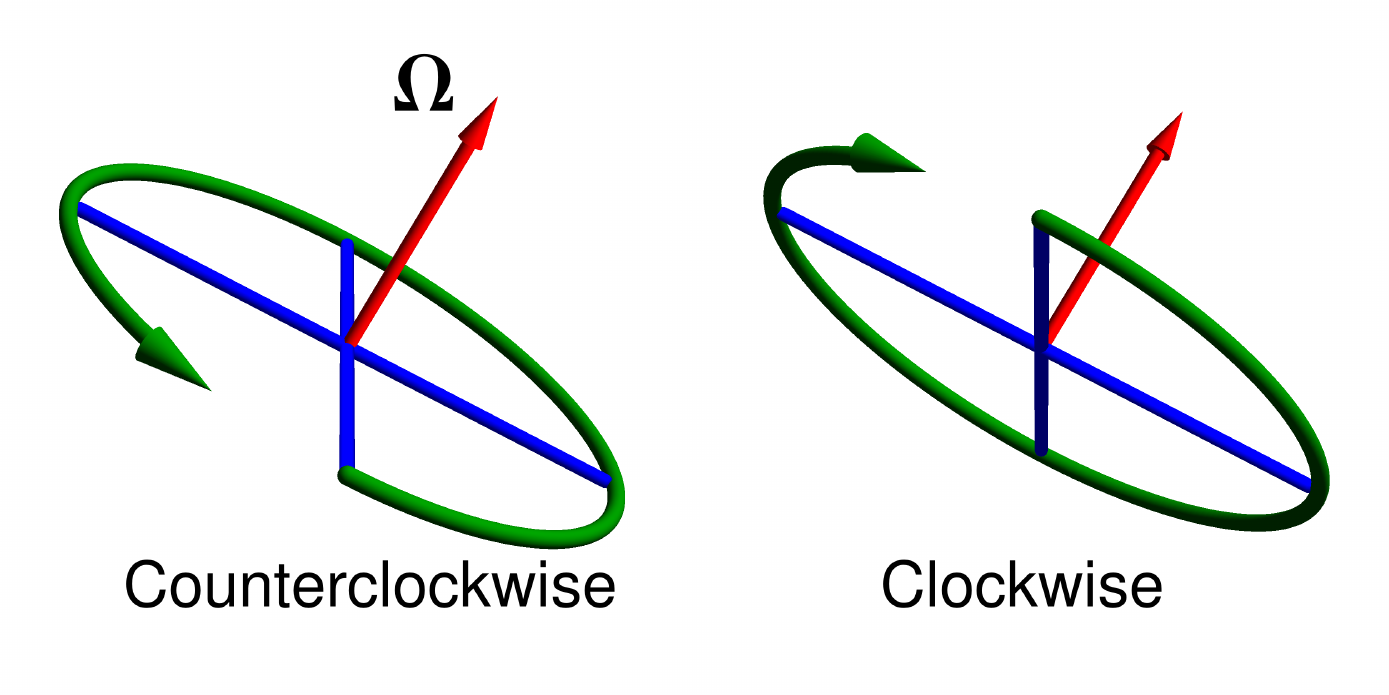}
\caption
{ 
Illustration of the degeneracy in supplementary inclinations of an accretion disk. For accretion disks with the two inclinations shown, the Doppler boosting will be identical (neglecting effects from black hole spin). As a result, the two images will have similar time-averaged appearances, but the direction of the flow on the sky will be opposite. An extreme example of this degeneracy is an accretion disk viewed face-on.
}
\label{fig_BasicGeometry}
\end{figure}

Past EHT observations of \sgra\ \citep{Doeleman_2008,Fish_2011} have already suggested structure that is more compact than the size of the photon ring that bounds the black hole ``shadow'' \citep{Bardeen_1973,Falcke_2000,Takahashi_2004}. This compact structure is most commonly explained via an accretion disk with its angular momentum axis inclined relative to the line of sight, which results in an image that is dominated by a small Doppler-boosted patch on the oncoming edge of the disk \citep[e.g.,][]{Dexter_2010,Broderick_2011,Psaltis_2015}. However, given that the current data are extremely sparse, the inclination and image properties cannot yet be confidently constrained. In addition, current estimates of the inclination are subject to a degeneracy between supplementary inclinations, $\{ \theta, \pi - \theta\}$, (see Figure~\ref{fig_BasicGeometry}) because simulated images of accretion flows exhibit a near symmetry orthogonal to the rotation axis \citep{Moscibrodzka_2009,Dexter_2010,Broderick_2011,
Shcherbakov_2012,Psaltis_2015}.

As ever more sophisticated models are fit to the data, it is equally important to develop model-independent assessments of the data. Chief among these is synthesis imaging \citep[see, e.g.,][]{Lu_2014,Fish_2014}. Yet, with a mass of $4\times 10^6\,M_{\odot}$ \citep{Ghez_2008,Gillessen_2009}, \sgra\ has a gravitational timescale of only $GM/c^3 \approx 20~{\rm seconds}$ and an orbital period at the innermost stable circular orbit of only $4{-}30$ minutes, depending on spin \citep{Bardeen_1972}. These short timescales suggest that conventional Earth-rotation synthesis imaging will be inapplicable for \sgra\ and, in the best case, will ignore the rich physics encoded within the variability, such as turbulence, orbital motion, and flaring. Thus, in contrast with past work to infer properties of the quiescent image using non-imaging EHT data products, our present emphasis is to study \emph{dynamics} of the emitting material with EHT data. 

Several authors have already explored how non-imaging VLBI can be applied to study rapid temporal variability. For instance \citet{Broderick_Loeb_2005,Broderick_Loeb_2006} simulated an orbiting ``hot spot'' around \sgra\ and calculated the expected images at submillimeter and near-infrared wavelengths. \citet{Doeleman_Hotspots} and \citet{Fish_Hotspots} then showed that EHT baselines can sensitively detect periodicities associated with these hot spots. Addressing more general circumstances, \citet{Broderick_2011} suggested that phase-referenced observations with the EHT may allow microarcsecond tracking of the image centroid on orbital timescales, and \citet{Johnson_2014} showed that polarimetric VLBI with the EHT is capable of microarcsecond astrometry of compact flaring structures, even for faint, non-periodic flares. 

In this paper, we explore a different metric: the temporal covariance between pairs of baselines. For baselines that are of a similar length and orientation, this covariance is a sensitive probe of image \emph{rotation}. Because realistic flows around black holes are subject to shearing and other secular evolution, these measurements can be used to determine orbital timescales using nearby ``snapshots'' between which the flow undergoes stable rotation. The temporal covariance then reflects the direction of the flow on the sky (clockwise or counter-clockwise), the angular velocity of the emitting material, and the radial distribution of emitting material. Estimating the direction of the flow immediately breaks the degeneracy in supplementary inclinations when analyzing quiescent images permitting an unambiguous determination of the angular momentum axis of the accretion flow. Importantly, our work requires no assumptions about the appearance of the quiescent structure or about precise flow dynamics. 
We only assume that there is an inhomogeneous, rotating component.

We begin, in \S\ref{sec::Accretion}, with a brief discussion of accretion and orbital dynamics near a black hole. Next, in \S\ref{sec::Interferometry}, we derive interferometric relationships for a differentially rotating flow. 
Then, in \S\ref{sec::GRMHD}, we apply our technique to synthetic observations of three-dimensional GRMHD simulations. 
In \S\ref{sec::Application_to_EHT}, we consider practical limitations when using this technique on EHT data products and discuss the EHT baseline pairs that are especially well-suited to this purpose. Finally, in \S\ref{sec::Summary}, we summarize our main results.

\section{The Accretion Flow Near a Black Hole}
\label{sec::Accretion}

The perceived angular velocity of a rotating accretion flow by a distant observer depends on properties of the spacetime, the accretion flow itself, and the viewing geometry. We now discuss each of these in turn. 

\subsection{Effects of the Spacetime on Orbital Velocities}
\label{sec::Spactime}
In the Kerr spacetime for a rotating black hole with dimensionless spin $0 \leq a \leq 1$ and gravitational radius $r_{\rm G} \equiv GM/c^2$, the period $P_{\rm orb}$ of a circular equatorial orbit at radius $r$ (in Boyer-Lindquist coordinates) can be written as \citep{Bardeen_1972}
\begin{align}
\label{eq::Porb}
P_{\rm orb}(r)=2\pi \left[\left(r/r_{\rm G} \right)^{3/2} \pm a \right] t_{\rm G},
\end{align}
where $+/-$ corresponds to prograde/retrograde orbits. 
For a Schwarzschild black hole ($a=0$), this expression reduces to $P_{\rm orb}(r)=2\pi \sqrt{r^3/(G M)}$, which is familiar as Kepler's third law. Because of the differential rotation implied by Eq.~\ref{eq::Porb}, large coherent features in a Keplerian flow will shear apart on an orbital timescale, quickly eliminating pure periodicities in light curves.\footnote{For instance, in a purely Keplerian flow $P_{\rm orb}\left( r(1+f) \right) \approx \left[ 1 + \frac{3}{2}f \right] P_{\rm orb}(r)$, so a feature situated at a radius $r$ and having a radial extent of $0.16 r$ will shear by approximately a quarter of an orbit each orbital period.}
For flows viewed off-axis, the effects of strong gravitational lensing can also change the apparent angular velocity depending on orbital phase.

Although no significant periodicities have been detected for \sgra, some simulations suggest that the radio emission is dominated by material at a Boyer-Lindquist radius of ${\sim}5r_{\rm G}$ \citep{Shiokawa_Thesis}. 
Because all emissions from small radii are lensed to similar apparent radii \citep[see, e.g.,][]{BNL_2009}, time-averaged images may not be able to constrain the emission radius as tightly as the orbital dynamics. However, orbital periodicities will probably be unable to meaningfully constrain the black hole spin, as noted by \citet{Broderick_Loeb_2005,Broderick_Loeb_2006}. For instance, at a radius of $5r_{\rm G}$, the effect of spin on the period of prograde orbits is ${\lsim}10\%$ and is degenerate with a $6\%$ change in emission radius. Thus, given the uncertainty of the precise emission radius at submillimeter wavelengths, the black-hole spin can probably not be securely estimated from its effect on orbital periodicities. 

However, while the orbital period at a given radius is rather insensitive to spin, the radius $r_{\rm ISCO}$ of the innermost stable circular orbit (ISCO) is not. For a maximally spinning black hole, $r_{\rm ISCO}$ ranges from $r_{\rm G}$ to $9r_{\rm G}$ for prograde and retrograde orbits, respectively. This steep dependence on spin was used by \citet{Doeleman_2012} to infer that the accretion disk in M87 is undergoing prograde rotation by associating the apparent size of the emission region with the lensed ISCO. Because measurements of the orbital period can accurately estimate the emission radius, these estimates could then provide a meaningful upper-bound on $r_{\rm ISCO}$ and, thus, a lower-bound on the signed spin ($+/-$ for prograde/retrograde). 

\subsection{Effects of the Accretion Flow on Orbital Velocities}

Because \sgra\ is highly under-luminous, emitting only ${\sim}10^{-9}$ of its Eddington luminosity, the accretion flow near \sgra\ is most likely a hot, thick disk \citep[e.g.,][]{Yuan_2003,Loeb_Waxman_2007,Yuan_2014}. Such disks tend to have sub-Keplerian rotation profiles because of their strong pressure support \citep{Narayan_Yi_1994,Yuan_2014}. The fractional reduction in angular velocity $\Omega$ relative to Keplerian $\Omega_{\rm K}$ is second order in the fractional scale height $H/R$ of the disk \citep[e.g.,][Eq.~20]{Blandford_Begelman_1999}. GRMHD simulations of \sgra\ typically have $H/R$ ranging from $0.1$ to $0.3$ \citep[e.g.,][]{Dexter_2010}, so the pressure support may decrease the angular velocities by ${\lsim}10\%$. For the particular GRMHD simulation used in this paper (see \S\ref{sec::GRMHD}), the average azimuthal fluid velocity differs from the Keplerian velocity by only ${\sim}1\%$ outside the ISCO. 

A potentially more serious limitation concerns the pattern velocities of emission features, which may differ from their underlying fluid velocities.\footnote{We thank the referee for identifying the importance of the emission pattern velocity.} For the GRMHD simulations that we discuss in \S\ref{sec::GRMHD}, the pattern velocity is 30-40\% lower than the fluid velocity at the radius of peak emission (${\approx}5r_{\rm G}$); the difference becomes even more pronounced closer to the event horizon. Although we are not aware of simulations that exhibit a predominantly counter-rotating submillimeter emission pattern, the relationship between the pattern and fluid velocities is an important consideration when interpreting measured angular velocities and will be analyzed in depth elsewhere (Shiokawa et al., in preparation).

\subsection{Effects of the Viewing Geometry on Orbital Velocities}
\label{sec::Effects_of_ViewingGeometry}

When rotation is not viewed face-on, apparent angular velocities depend on orbital phase. For instance, consider a point particle in a circular orbit at an inclination $0\leq \theta \leq \pi/2$ relative to the line of sight (i.e., the motion of a bead on a circular wire). At an orbital phase $\phi$, the apparent angular velocity is 
\begin{align}
\label{eq::Projected_Omega}
\Omega(\phi;\theta) \propto \frac{\cos\theta}{\cos^2\theta + \cos^2 \phi \sin^2\theta}.
\end{align}
At $\phi=0$ the particle is moving away from the observer, and at $\phi=\pi$ is moving toward the observer. The velocity component along the line of sight is $\cos\phi \sin \theta$. Note that the velocity is zero for an edge-on view of the orbit ($\theta=\pi/2$) because in that case the motion is restricted to a line and has no angular velocity. Near a black hole, relativistic aberration and lensing will also affect the apparent velocity. For instance, emission will be lensed above and below the black hole, producing a non-zero apparent angular velocity even for the edge-on case. 

The orbit-averaged angular velocity must, of course, agree with the true angular velocity. However, because of Doppler effects and lensing, the emitting material will vary in brightness through the orbit. For instance, letting $\mathcal{D} \equiv \gamma^{-1}(1-\beta_{\parallel})^{-1}$ denote the Doppler factor, where $\gamma = (1-\beta^2)^{-1/2}$ is the Lorentz factor and $\beta$ is the normalized velocity in the rest frame of the observer, the observed flux density is scaled by a factor of $\mathcal{D}^{3+\alpha}$ relative to that in the co-moving frame, where $\alpha$ is the spectral index of the emission \citep[e.g.,][]{Blandford_Konigl_1979}. Consequently, the orbit-averaged angular velocity weighted by $\mathcal{D}^3$ will underestimate the true angular velocity; most emission comes when the orbit is approaching the observer, with a low apparent angular velocity. 

An additional complication arises because the angular velocity must be defined relative to a particular centroid -- typically the centroid of the quiescent flux. For an inclined flow, because of Doppler boosting on the approaching side of an accretion disk, this centroid will not be centered on the black hole. We will now derive a strategy for estimating the angular velocity using interferometric visibilities that mitigates this effect.

\section{Lagged Interferometric Covariance}
\label{sec::Interferometry}

\subsection{Interferometric Observations of a Rotating Flow}
\label{sec::Interferometry_sub}

We now explore how signatures of a rotating flow are manifest in interferometric observables. The interferometric visibility $\tilde{I}(\textbf{u})$ measured by a baseline $\textbf{u}$ is related to the source brightness distribution $I(\textbf{x})$ via the van~Cittert-Zernike Theorem \citep{TMS}:
\begin{align}
\label{eq::vCZ}
\tilde{I}(\textbf{u},t) &= \int d^2\textbf{x}\, I(\textbf{x},t) e^{-2\pi i \textbf{u} \cdot \textbf{x}}. 
\end{align}
In this expression, $\textbf{u}$ is the vector baseline orthogonal to the line of sight, in wavelengths, and $\textbf{x}$ is an angular coordinate on the sky, in radians. We have included a time coordinate, $t$, to account for the possibility of a changing source emission structure with time. 

From Eq.~\ref{eq::vCZ}, it is evident that an image rotation by some angle $\theta$ about the origin $\textbf{x}=\textbf{0}$ leads to an identical rotation in the visibility domain about $\textbf{u}=\textbf{0}$.  
Moreover, standard interferometric observables  -- visibility amplitudes and closure phases -- are unaffected by a shift of the image center: $\textbf{x} \rightarrow \textbf{x} + \Delta \textbf{x}_0$. Consequently, these observables for one set of baselines sampling the unrotated image will be equivalent for the same baselines rotated by $\theta$ but sampling the image after it is rotated by $\theta$ about \emph{any} fixed point $\textbf{x}_0$. This property helps to mitigate the angular-velocity bias from an offset image center (\S\ref{sec::Effects_of_ViewingGeometry}) when studying angular velocities.

Hence, as long as the image is not azimuthally symmetric, stable rotating structures will introduce a lagged correlation between visibilities on pairs of baselines with similar lengths but different orientations. The angular velocity of the rotating flow is then given by the angular difference between the baseline directions divided by the temporal lag corresponding to the peak covariance. This inference, which determines both the direction and angular velocity of rotation, is determined entirely in the visibility domain and can be achieved with as few as two baselines (three stations).

\begin{figure*}[th]
\centering
\includegraphics*[width=0.77\textwidth]{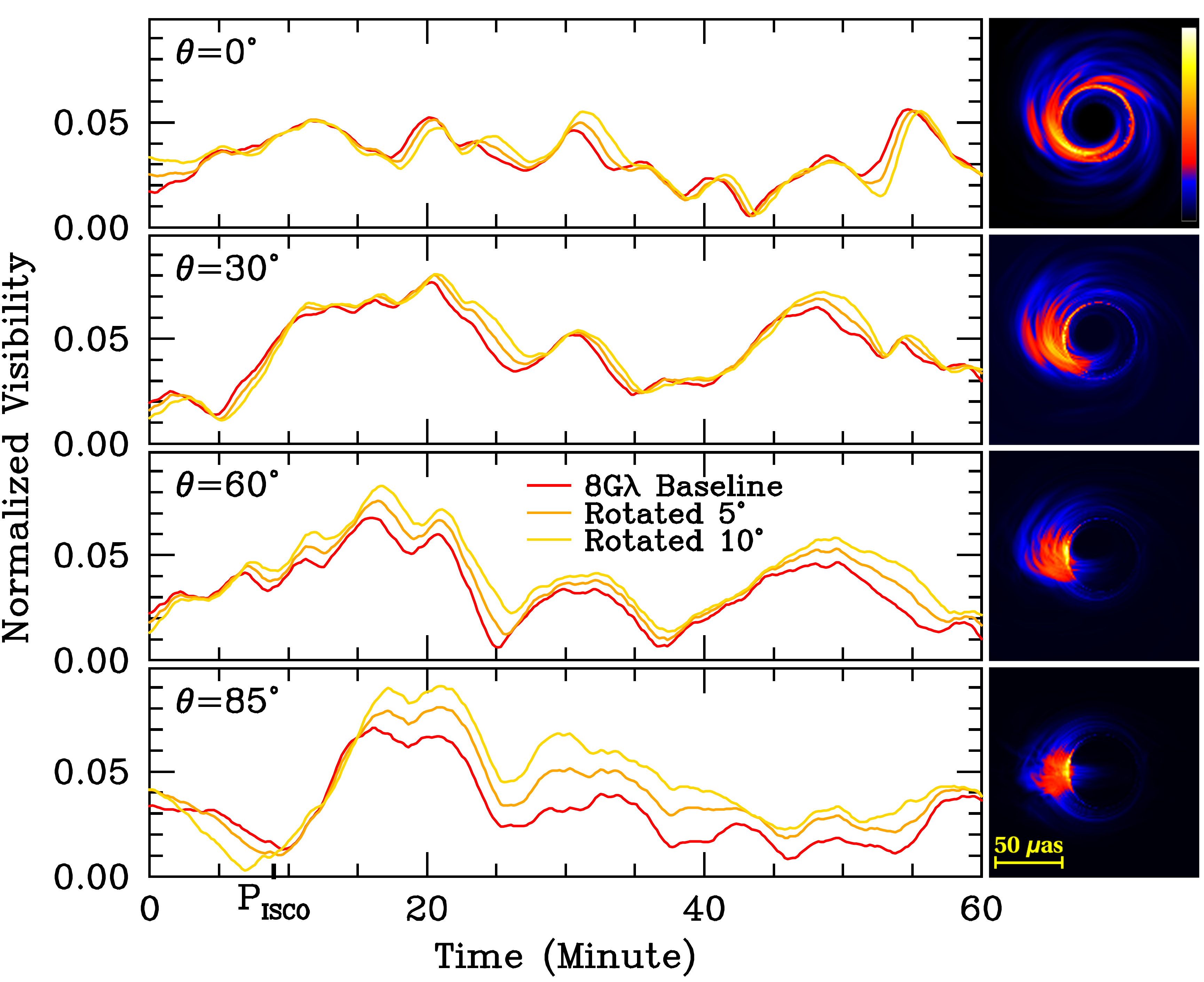}
\caption
{ 
Left panels show normalized visibility as a function of time for a simulated GRMHD movie sampled on three baselines of identical lengths ($8~{\rm G}\lambda$) but slightly different orientations. Results are shown for four viewing inclinations: $\theta=0^{\circ}$ (face-on), $\theta=30^{\circ}$, $\theta=60^{\circ}$, and $\theta=85^{\circ}$. 
Corresponding images on the right show single frames for each case, with a linear color scale. For these images and visibilities, we show results before accounting for interstellar scattering (see \S\ref{sec::Scattering}). 
At low inclinations, the lags between the time series are readily apparent by comparing the turning points of each curve and arise because of rotation of the irregular flow. Especially for the face-on disk, these lags accurately estimate the orbital periodicity, even though the visibility curves show no clear periodicities. In contrast, for the nearly edge-on case ($\theta=85^\circ$), the lags are irregular in both magnitude and sign.
}
\label{fig_SampleTimeseries}
\end{figure*}

\vspace{0.1cm}

\subsection{Identifying the Peak Lagged Correlation}
\label{sec::Peak_Lag}

We now describe our procedure to estimate the lagged correlation. To properly identify the peak lag, we must address three potential sources of contamination: 1.)~There will be overall changes in the total flux of the image (i.e., the zero-baseline visibility), 2.)~There will be slow secular evolution of the bulk emission structure and of observing parameters (e.g., from rotation of the Earth), and 3.)~The brightness centroid may not be centered on the black hole. 

The first consideration is especially important for short baselines, where variations in the interferometric visibility will be tightly correlated with modulation of the total image flux. As a result, lagged correlations will have a peak at zero lag. To eliminate this feature and to account for source flux modulation, it is advantageous to work with normalized visibilities -- i.e., visibilities divided by the simultaneous zero-baseline visibility. To mitigate the second contaminating effect, we subdivide each long time-series into shorter segments and determine the peak lag for each separately. This subdivision also naturally accommodates observational constraints, such as regular breaks for calibration or pointing scans. Finally, we eliminate the third source of contamination by studying only the normalized visibility magnitudes, as discussed in \S\ref{sec::Interferometry_sub}.

In each segment, we then estimate the lagged cross-correlation $\rho(\Delta t)$ using the classical estimator:
\begin{align}
\label{eq::rho}
\rho(\Delta t) &\equiv \frac{ \left \langle \left[ A(t) - \langle A(t) \rangle \right] \left[ B(t{+}\Delta t) - \langle B(t{+}\Delta t) \rangle \right] \right \rangle   }{\sigma_A \sigma_B}, 
\end{align}
where $\{A(t),B(t)\}$ are the normalized visibilities $\left|\tilde{I}(\textbf{u},t)/\tilde{I}(\textbf{0},t) \right|$ on the pair of baselines, and $\sigma_x$ denotes the standard deviation of the time series $x$. 

When the EHT is complete and begins collecting regular data on \sgra, additional knowledge of the variability can be applied to develop more sophisticated estimators of the lagged covariance. 
For example, differences of nearby measurements (approximating the time-series' derivatives) are effective for de-trending and whitening stochastic time series \citep{Brockwell_Davis,Box_Jenkins}, and may facilitate superior estimates of the lag. 
Also, alternative metrics such as the Discrete Correlation Function \citep{Edelson_Krolik_1988} could be adopted for unevenly sampled data, such as from irregularly interspersed scans on calibration targets, although VLBI scans can be correlated at arbitrarily short segmentation times. 
However, since our focus is merely a proof-of-concept, we will use the simple correlation estimate defined by Eq.~\ref{eq::rho}.

\section{Application to GRMHD Simulations}
\label{sec::GRMHD}

We tested our new technique by applying it to a 3D GRMHD simulation \citep[\texttt{b0-high} from][]{Shiokawa_Thesis} of a radiatively inefficient accretion flow \citep[e.g.,][]{Esin_1996} onto a massive ($4.5{\times}10^6~M_\odot$) spinning black hole \citep{Gammie_2003,Noble_2006}. We started the simulation with a hot, geometrically thick, and tenuous disk \citep{Fishbone_Moncrief_1976} around the black hole.  The disk was seeded by a weak poloidal magnetic field so that the magnetorotational instability (MRI) could grow, driving accretion.  The radii of the initial inner edge and pressure maximum of the disk were $12r_{\rm G}$ and $24r_{\rm G}$, respectively.  We set the dimensionless spin parameter of
the black hole to be $a=0.9375$, following the ``best-bet model" for \sgra\ from \citet{Moscibrodzka_2009}. For this spin, the ISCO radius is $r_{\rm ISCO}=2.04 r_G$ with a corresponding orbital period of $P_{\rm ISCO}=24.25t_{\rm G}=8.96~{\rm minutes}$, where $t_{\rm G}=G M/c^3=22.17$ seconds. However, most of the 230~GHz emission originates from material at a Boyer-Lindquist radius of $6-5r_{\rm G}$ for viewing inclination of $0^\circ-80^\circ$, respectively, with a corresponding orbital period of $P_{\rm orb}\sim 30~{\rm minutes}$ \citep{Shiokawa_Thesis}. Note that spin has a ${\lsim}10\%$ effect on orbital period at this radius and amounts to at most a factor of 2 even at $r = r_{\rm G}$.  

Our $260{\times}192{\times}128$ simulation grid was defined by modified spherical coordinates: logarithmically
scaled radial coordinates spanning $1.22-240r_{\rm G}$, poloidal coordinates with $2^\circ$ cutouts at the poles to avoid the
coordinate singularities, and azimuthal coordinates spanning the full $2\pi$. The MRI saturated at the initial pressure maximum
radius around $t\sim 8000t_{\rm G}$; we then ran the simulation for an additional $6500t_{\rm G}$, which defined the data used in our subsequent analysis.

\begin{figure*}[t]
\centering
\includegraphics*[width=1.0\textwidth]{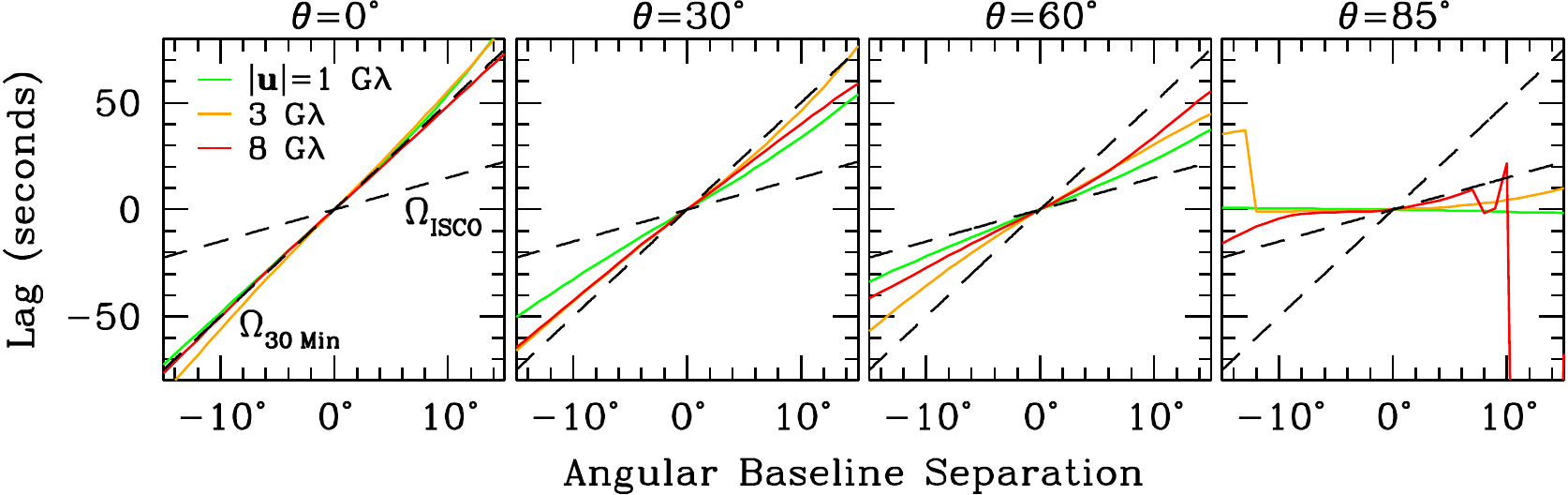}
\caption
{ 
Peak lag as a function of baseline separation for four inclinations: $\theta=0^{\circ}$ (face-on), $\theta=30^{\circ}$, $\theta=60^{\circ}$, and $\theta=85^{\circ}$ (see Figure~\ref{fig_SampleTimeseries} for characteristic images). In each case, results are shown for three baseline lengths, $1~{\rm G}\lambda$,  $3~{\rm G}\lambda$, and $8~{\rm G}\lambda$ (respective resolutions of $\approx 200~\mu{\rm as}$, $70~\mu{\rm as}$, and $25~\mu{\rm as}$). 
The panels also indicate the angular-velocity curves for material in circular orbits at the ISCO and for a 30-minute orbital period, which is close to the period for material that dominates the emission ($r \sim 5r_{\rm G}$). For inclinations up to $60^\circ$, the derived direction of the flow is correct, but long baselines are increasingly important at higher inclination to measure a meaningful angular velocity, and an upward bias in the inferred rotational velocities is important for $\theta \gsim 60^{\circ}$ because of the inclination dependence of projected angular velocity (see, e.g., Eq.~\ref{eq::Projected_Omega}). For an inclination of $85^\circ$ (close to edge-on), our method cannot identify a predominant rotation direction (by symmetry), so it produces irregular results that will vary among different baseline pairs and observing epochs.
}
\label{fig_VaryingInclination}
\end{figure*}

For the radiative transfer, we performed general relativistic ray-tracing by
integrating synchrotron emission and absorption along photons' geodesics until
they escape the simulation box and fall into each pixel of the ``camera"
\citep{Noble_2007}. 
Unlike most previous studies, our radiative transfer does not use the ``fast light'' approximation, which assumes that all the photons emitted in one time slice arrive simultaneously at the camera. We instead account for evolution of the fluid as each photon propagates. 
We assumed the electron distribution function to be
thermal and the proton-to-electron temperature ratio to be 3. At $\lambda{=}1.3~{\rm mm}$, contributions
from bremsstrahlung and Compton scattering are negligible.  Because the disk mass can be chosen arbitrarily
in the conversion from simulation units to physical units (the disk evolution is independent of its mass in the regime where self-gravity and radiative effects are negligible), we chose a value so that the simulation's time-averaged flux density was comparable to observed values for \sgra\ at $\lambda{=}1.3~{\rm mm}$ \citep[e.g.,][]{Bower_2015}.

To test our proposed method, we used 1700 frames spaced by $0.5t_{\rm G}\approx 11$ seconds, equivalent to a 5.2-hour
observation. Figure~\ref{fig_SampleTimeseries} shows example image snapshots and time series for interferometric visibilities at three different viewing inclinations. Note that these time series do not reflect the orbital periodicities, as was also noted by \citet{Dolence_2012} for total-flux light curves at $\lambda{=}1.3~{\rm mm}$ from similar simulations. 

To account for slow trends in the data, we divided each time series into 200-frame segments (37 minutes) and averaged the peak lags calculated
separately in each, as discussed in \S\ref{sec::Peak_Lag}. Figure~\ref{fig_VaryingInclination} shows the resulting peak lag as a function of angular baseline separation for three baseline lengths and for three viewing inclinations. When the accretion flow is viewed face-on, the inferred orbital period is close to the value for material at the radius of maximum emission ($P_{\rm orb} \approx 30~{\rm minutes}$). At an inclination of $\theta=30^\circ$, the inferred orbital period is only accurate for long baselines (${\gsim}3~{\rm G}\lambda$).  At $\theta=60^\circ$, the inferred orbital periods are lower than the true value by a factor of ${\sim}2$, even on long baselines, but the inferred direction of the flow on the sky is correct in every case. At $\theta=85^\circ$, the peak lag varies erratically in sign and magnitude, as is expected from the near symmetry of the image in this case. 

Note that when the inferred periodicity is stable (i.e., at low inclinations) it is faster than the orbital period at the ISCO for a non-rotating black hole (in this example, 34 minutes), so these measurements could be used to argue that the black hole spin was non-zero and that the accretion flow was in a prograde orbit. In practice, to ensure a robust measurement of angular velocity and direction will require confirmation at different observing epochs and on different baseline pairs, ideally sampling different position angles. Variations among different baseline pairs and different observing epochs would indicate that the inferred angular velocity and direction are not meaningful and could provide evidence that the flow is being viewed at high inclination.

\section{Considerations for the EHT}
\label{sec::Application_to_EHT}

\subsection{Ideal EHT Baselines}
\label{sec::Ideal_EHT_Baselines}

\begin{figure}[t]
\centering
\hspace{1.4cm}\includegraphics*[height=0.34\textwidth]{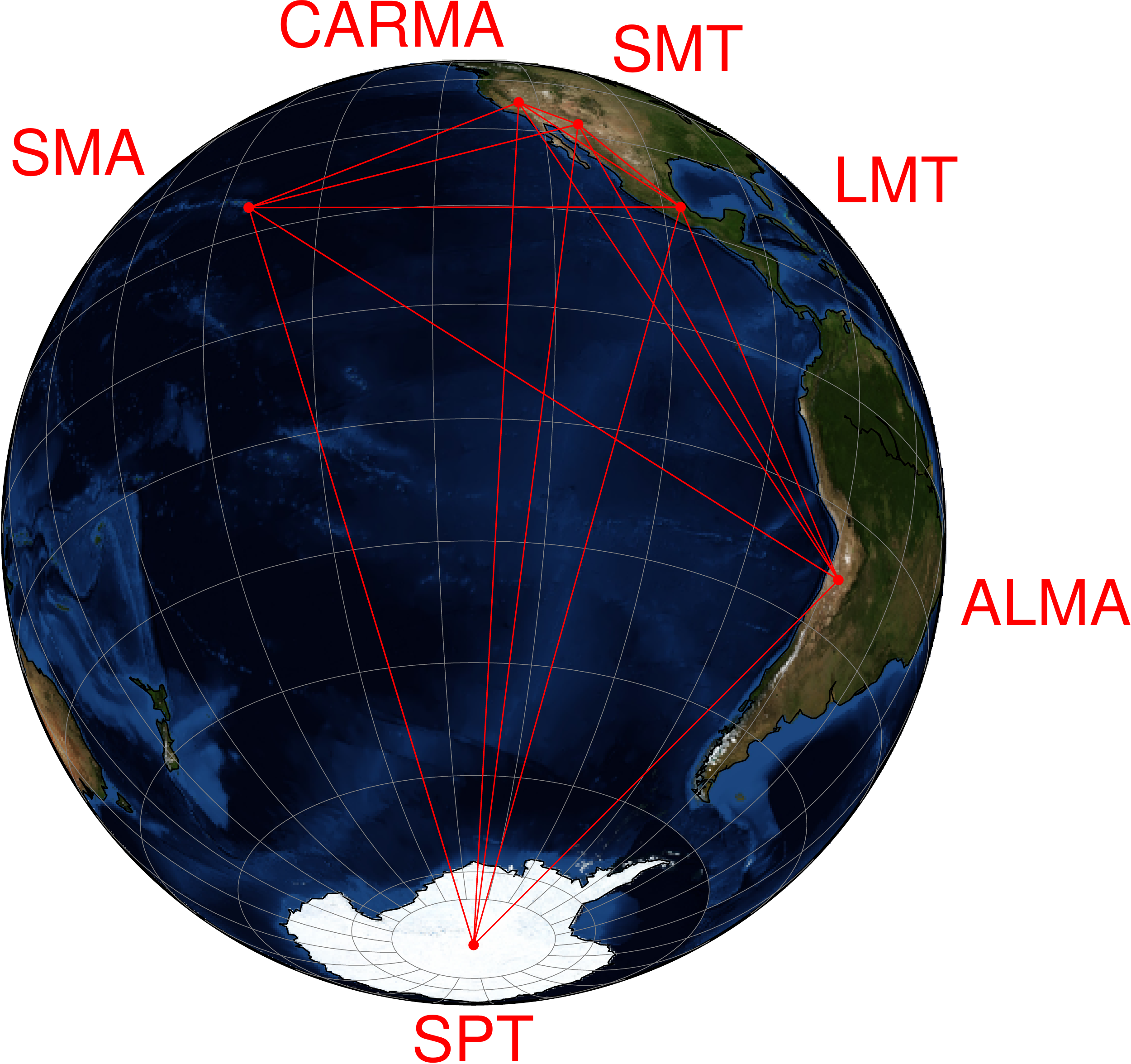}\\ \vspace{0.2cm}
\includegraphics*[height=0.35\textwidth]{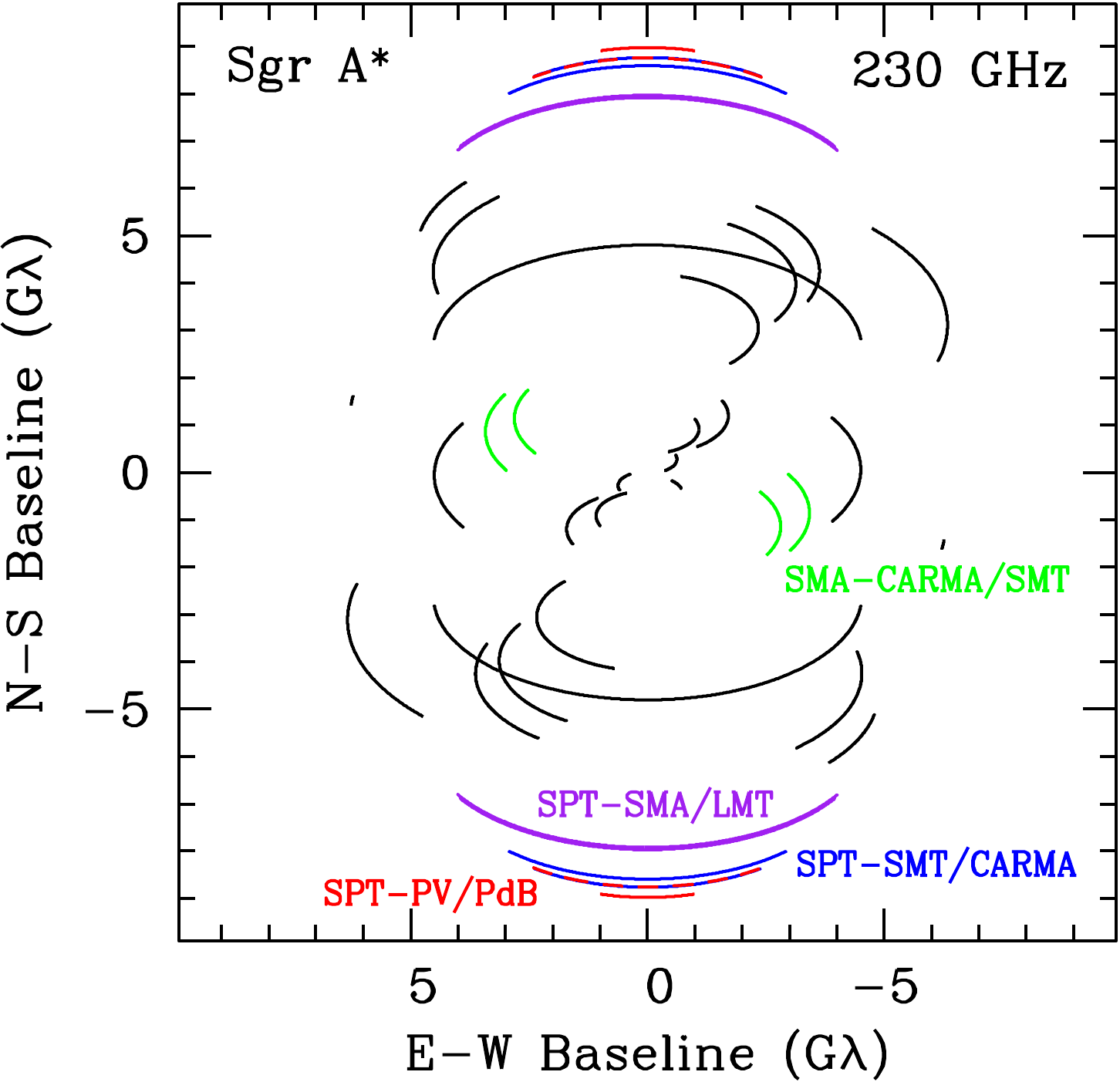}\hspace{0.5cm}
\caption
{ 
({\it Top}) The planned EHT array as seen from the declination of \sgra; PV and PdB are not visible. ({\it Bottom}) Corresponding baseline tracks for all planned EHT sites. Colored tracks show the candidate baseline pairs listed in Table~\ref{tab::Baseline_Pairs}. Baseline lengths are given in gigawavelengths at $\lambda=1.3~{\rm mm}$. Note that because the latitude of PV ($37.1^{\circ}$) is nearly identical to CARMA ($37.3^\circ$), their baselines to the SPT are nearly identical. 
}
\label{fig_EHT}
\end{figure}

The EHT has several promising baselines to study lagged correlation for \sgra\ (see Figure \ref{fig_EHT}); baselines from pairs of sites at similar latitude to the South Pole Telescope are especially well-suited to our method. For instance, the pair of baselines from the South Pole Telescope (SPT) to the Submillimeter Array (SMA) and to the Large Millimeter Telescope (LMT) have a mutual visibility of approximately 4 hours, assuming a $15^\circ$ elevation pointing limit. Over this entire span, the ${\sim} 8~{\rm G}\lambda$ projected baseline lengths differ by no more than 1\%, while the angular difference between the baselines ranges from $29.4^\circ$ to $32.8^\circ$. A 30-minute orbital periodicity would have a corresponding peak lag of ${\approx} 1~{\rm minute}$ between these baselines. 

Even the current EHT may be suitable for this method. The baselines from the SMA to SMT and CARMA have lengths (${\sim} 3~{\rm G}\lambda$) with ratios between $0.8-0.9$ for their 4 hours of mutual visibility and a baseline rotation of $5.8-9.2^\circ$.\footnote{The CARMA observatory was shut down following the 2015 EHT campaign. However, the CARMA site is still relevant for continued observations with the EHT because a nearby site may be added (associated with the Owens Valley Radio Observatory).} Table \ref{tab::Baseline_Pairs} provides details for four of the most promising baseline pairs.

{
\begin{deluxetable*}{lccccc}
\tablewidth{\textwidth}
\tablecaption{Exemplar Baseline Pairs for the EHT.}
\tablehead{ 
\colhead{ Baseline Pair } & \colhead{ Mutual Visibility (hours) } & \colhead{ Resolution ($\mu{\rm as}$) } & \colhead{ Max Length Ratio } & \colhead{ Angular Difference } & \colhead{ Lag$\times \frac{P_{\rm orbit}}{30~{\rm minutes}}$ (seconds) }   }
\startdata
SPT-SMA/LMT        & 4.0 & 25 & 1.01 & $29.4^\circ-32.8^{\circ}$ & $74-82$ \\
SPT-SMT/CARMA      & 5.0 & 25 & 1.02 & $2.2^\circ-4.4^{\circ}$   & $6-11$ \\
SPT-PV/PdB         & 2.2 & 25 & 1.03 & $3.2^\circ-5.0^{\circ}$   & $8-12$ \\
SMA-CARMA/SMT      & 3.9 & 65 & 1.22 & $5.8^\circ-9.2^{\circ}$   & $14-23$ 
\enddata
\label{tab::Baseline_Pairs}
\tablecomments{\emph{SPT:} South Pole Telescope, \emph{SMA:} Submillimeter Array, \emph{CARMA:} Combined Array for Research in Millimeter-wave Astronomy, \emph{SMT:} Submillimeter Telescope, \emph{LMT:} Large Millimeter Telescope, \emph{ALMA:} Atacama Large Millimeter/submillimeter Array, \emph{PV:} Institut de Radioastronomie Millim\'{e}trique (IRAM) telescope on Pico Veleta, \emph{PdB:} IRAM Plateau de Bure Interferometer.}
\end{deluxetable*}
}

\subsection{Temporal Resolution of the EHT}

Because VLBI data can be correlated and sampled on arbitrary segmentation times, temporal resolution is not likely to be a fundamental limitation for our proposed technique. The EHT will even achieve signal-to-noise ratio ${\gsim}1$ on scans lasting only seconds or less. For example, consider the SPT-SMA baseline. This baseline will have an effective system equivalent flux density (SEFD) of ${\sim}5000~{\rm Jy}$. Then, for 4~GHz of bandwidth, a 10-second scan would have a thermal noise of $\sim 20~{\rm mJy}$. In the simulations discussed in \S\ref{sec::GRMHD}, the source flux ranges from 10 to 200 mJy on this baseline. Even after accounting for interstellar scattering, these values will only be reduced by a factor of ${\sim}2$ (see \S\ref{sec::Scattering}). Thus, the signal-to-noise ratio may be greater than unity even on 10-second scans. See \citet{Lu_2014} for a list of current EHT SEFDs.

\subsection{Effects from Calibration Uncertainties}

Precise calibration poses a major challenge for interferometry and may complicate our proposed method. Although our method does not require phase information, amplitude calibration must still be stable to within the variability amplitude to avoid contaminating the covariance. This limitation may render long baselines most useful, where fractional variations of the signal are likely larger.

Alternatively, one can eliminate station-based calibration errors using ``closure'' quantities \citep{TMS}. Because the most common closure quantities -- closure phase and closure amplitude -- involve multiple baselines, they are not directly useful for our purposes. However, fractional polarization is baseline-based and provides identical immunity after calibrating slowly changing differential gain and leakage terms \citep[e.g.,][]{RWB94}. Comparisons with simulations that include polarization information \citep[e.g.,][]{Shcherbakov_2012,Shcherbakov_2013,Dexter_2014} will allow us to assess whether lagged correlation of fractional polarization on close baseline pairs can likewise reflect dynamics of the bulk flow. However, because the polarization direction can change throughout an orbit, from the changing local magnetic field direction or relativistic aberration or from strong-field relativistic effects such as lensing or parallel transport, the application to polarization may require significant 
modification.

\subsection{Effects from Interstellar Scattering}
\label{sec::Scattering}
Propagation through the turbulent interstellar medium scatters radio waves and causes wavelength-dependent blurring of images. The scattering is especially strong along the line of sight to the Galactic Center, and although interstellar scattering is subdominant to the intrinsic structure at $\lambda=1.3~{\rm mm}$, it remains an important consideration for the EHT \cite[see][]{Fish_2014}. Nevertheless, interstellar scattering is not likely to be an important consideration for our proposed methods.  

For instance, the dominant effect of scattering, blurring via a deterministic image convolution, is invertible and simply decreases the signal-to-noise ratio on long EHT baselines by a factor of up to ${\sim}4$. Because the scattering is weaker in the North-South direction, long baselines to the SPT are only attenuated by a factor of ${\sim}2$. Regardless of baseline, this ensemble-average scattering effect will not bias the lagged covariance. 
 
In contrast, the subdominant effect of scattering, ``refractive noise,'' is not deterministic and will affect long-baseline properties \citep{NarayanGoodman89,GoodmanNarayan89,Johnson_Gwinn_2015}. However, refractive noise is persistent (changing over a timescale of ${\gsim}1~{\rm day}$) and wideband, so also will not affect the covariance or dynamical imaging that we propose, which rely on variability timescales of minutes.

\section{Summary}
\label{sec::Summary}

We have showed that the covariance between pairs of interferometric baselines with similar lengths and close angular separation can sensitively probe the angular velocity of emission for a rotating flow. This non-imaging technique can estimate both the direction and angular velocity of the flow on the sky with as few as two baselines (three stations). By employing baselines of close angular separation, one can accurately estimate orbital periods even if the rotating structures evolve significantly over a single orbit, as is expected from differential rotation in a Keplerian flow. 

Our primary motivation has been EHT observations of \sgra. While our proposed technique would be most effective for a face-on viewing geometry, which is disfavored by current VLBI constraints, we show that EHT baseline pairs can robustly estimate the direction and can roughly estimate the angular velocity even for a moderately-inclined rotation axis. In particular, a measurement of the orbital direction would break the degeneracy in supplementary inclinations, allowing unambiguous comparison with larger-scale features such as the circumnuclear disk, the inner stellar disk \citep[e.g.,][]{Bartko_2009}, and the more recently discovered G1 and G2 gas clouds on trajectories passing within the Bondi radius of \sgra\ \citep{Gillessen_2012,Pfuhl_2015,McCourt_Madigan_2015}. 

Despite the generality of our approach, there are significant remaining uncertainties that can affect the interpretation of inferred angular velocities. At high inclinations, the estimated angular velocities can be significantly biased (see Figure~\ref{fig_VaryingInclination}), and so the applicability to \sgra, which does not have a firmly established inclination, is not yet secure. Indeed, there is not yet a consensus on if the emission from \sgra\ arises in an accretion disk or a jet. Even for a disk viewed face-on, there may be significant differences between the pattern velocity of emission features and the fluid velocity. For a thick accretion disk, strong pressure support may affect rotational periods as well, invalidating direct comparisons with rotation curves in the Kerr metric. To resolve these remaining questions will require both observational input, which is imminent with the addition of many new EHT sites, and improved understanding of the relationship between accretion and emission properties in GRMHD simulations.

Although we have focused on analyzing emission from a relatively steady accretion flow, our approach can also be applied to emission dominated by flaring components in the accretion disk \citep[e.g.,][]{Broderick_Loeb_2006} or to rapid helical motion in a jet \citep[e.g.,][]{Broderick_Loeb_2009}. Such applications would provide valuable counterparts to potential astrometry of the flaring region with polarimetric VLBI \citep{Johnson_2014} or with near-infrared interferometry \citep[e.g.,][]{Hamaus_2009,Vincent_2011}. Our method may also be valuable for other observations of time-variable structures with sparse visibility data. For example, space-VLBI experiments, such as RadioAstron \citep{Kardashev_2013} and the planned mission Millimetron \citep{Wild_2009} will provide many baseline pairs with nearly identical length and orientation from a single space dish to ground stations. 

Moving beyond a simple lagged correlation, rapid rotation of the image could potentially be used to {improve} image reconstructions \citep[e.g.,][]{Sault_1997}. Namely, conventional imaging algorithms assume that the source is static throughout rotation of the Earth and then use the Earth's rotation to increase sampling of the unrotated image in the visibility domain (termed Earth-rotation synthesis imaging). For the accretion flow of \sgra, the situation is reversed: the Earth is nearly static for an entire rotation of the source. This correspondence suggests that 
\emph{source}-rotation synthesis imaging may allow an array to achieve better effective visibility coverage on timescales of {minutes} than would be possible with a full night observing a static source. Thus, the EHT may be capable of rapid snapshot images of an inhomogeneous and rapidly-rotating flow, if the rotation curve of the source is well-understood.

\acknowledgments
We are pleased to acknowledge Charles Gammie for providing computation resources for the simulation, Scott Noble for originally developing HARM3D and the radiative transfer code we used, Olek Sadowski for discussions about orbital dynamics of thick accretion disks, and Laura Vertatschitsch for providing the Python code to generate Figure~\ref{fig_EHT}. We thank the referee for a thorough, insightful review that clarified the applicability and scope of our method. We thank the National Science Foundation (AST-1310896, AST-1312034, AST-1211539, and AST-1440254) and the Gordon and Betty Moore Foundation (\#GBMF-3561) for financial support of this work. \ \\

\bibliography{Baseline_Covariance.bib}

\begin{thebibliography}{69}
\expandafter\ifx\csname natexlab\endcsname\relax\def\natexlab#1{#1}\fi

\bibitem[{{Aitken} {et~al.}(2000){Aitken}, {Greaves}, {Chrysostomou},
  {Jenness}, {Holland}, {Hough}, {Pierce-Price}, \& {Richer}}]{Aitken_2000}
{Aitken}, D.~K., {Greaves}, J., {Chrysostomou}, A., {Jenness}, T., {Holland},
  W., {Hough}, J.~H., {Pierce-Price}, D., \& {Richer}, J. 2000, \apjl, 534,
  L173

\bibitem[{{Baganoff} {et~al.}(2003){Baganoff}, {Maeda}, {Morris}, {Bautz},
  {Brandt}, {Cui}, {Doty}, {Feigelson}, {Garmire}, {Pravdo}, {Ricker}, \&
  {Townsley}}]{Baganoff_2003}
{Baganoff}, F.~K., {et~al.} 2003, \apj, 591, 891

\bibitem[{{Bardeen}(1973)}]{Bardeen_1973}
{Bardeen}, J.~M. 1973, in Black Holes (Les Astres Occlus), ed. C.~{Dewitt} \&
  B.~S. {Dewitt}, 215--239

\bibitem[{{Bardeen} {et~al.}(1972){Bardeen}, {Press}, \&
  {Teukolsky}}]{Bardeen_1972}
{Bardeen}, J.~M., {Press}, W.~H., \& {Teukolsky}, S.~A. 1972, \apj, 178, 347

\bibitem[{{Bartko} {et~al.}(2009){Bartko}, {Perrin}, {Brandner}, {Straubmeier},
  {Richichi}, {Gillessen}, {Paumard}, {Hippler}, {Eckart}, {Sch{\"o}ller},
  {Eisenhauer}, {Haubois}, {Lenzen}, {Rabien}, {Cl{\'e}net}, {Ramos}, {Thiel},
  {Berger}, {Baumeister}, {Kellner}, {Cassaing}, {B{\"o}hm}, {Hofmann},
  {Gendron}, {Klein}, {Dodds-Eden}, {Houairi}, {Hormuth}, {Gr{\"a}ter},
  {Kervella}, {Naranjo}, {Genzel}, {F{\'e}dou}, {Henning}, {Hamaus}, {Jocou},
  {Neumann}, {Haug}, {Lacour}, {Laun}, {Kolmeder}, {Malbet}, {Rohloff},
  {Pfuhl}, {Perraut}, {Ziegleder}, {Rouan}, {Rousset}, {Amorim}, \&
  {Lima}}]{Bartko_2009}
{Bartko}, H., {et~al.} 2009, NewAR, 53, 301

\bibitem[{{Blandford} \& {Begelman}(1999)}]{Blandford_Begelman_1999}
{Blandford}, R.~D., \& {Begelman}, M.~C. 1999, \mnras, 303, L1

\bibitem[{{Blandford} \& {K{\"o}nigl}(1979)}]{Blandford_Konigl_1979}
{Blandford}, R.~D., \& {K{\"o}nigl}, A. 1979, \apj, 232, 34

\bibitem[{{Bower} {et~al.}(2006){Bower}, {Goss}, {Falcke}, {Backer}, \&
  {Lithwick}}]{Bower_2006}
{Bower}, G.~C., {Goss}, W.~M., {Falcke}, H., {Backer}, D.~C., \& {Lithwick}, Y.
  2006, \apjl, 648, L127

\bibitem[{{Bower} {et~al.}(2003){Bower}, {Wright}, {Falcke}, \&
  {Backer}}]{Bower_2003}
{Bower}, G.~C., {Wright}, M.~C.~H., {Falcke}, H., \& {Backer}, D.~C. 2003,
  \apj, 588, 331

\bibitem[{{Bower} {et~al.}(2014){Bower}, {Markoff}, {Brunthaler}, {Law},
  {Falcke}, {Maitra}, {Clavel}, {Goldwurm}, {Morris}, {Witzel}, {Meyer}, \&
  {Ghez}}]{Bower_2014}
{Bower}, G.~C., {et~al.} 2014, \apj, 790, 1

\bibitem[{{Bower} {et~al.}(2015){Bower}, {Markoff}, {Dexter}, {Gurwell},
  {Moran}, {Brunthaler}, {Falcke}, {Fragile}, {Maitra}, {Marrone}, {Peck},
  {Rushton}, \& {Wright}}]{Bower_2015}
---. 2015, \apj, 802, 69

\bibitem[{Box {et~al.}(2008)Box, Jenkins, \& Reinsel}]{Box_Jenkins}
Box, G.~E.~P., Jenkins, G.~M., \& Reinsel, G.~C. 2008, {Time Series Analysis:
  Forecasting and Control}, 4th edn., {Wiley Series in Probability and
  Statistics} (Hoboken, N.J.: John Wiley \& Sons, Inc.)

\bibitem[{Brockwell \& Davis(2002)}]{Brockwell_Davis}
Brockwell, P.~J., \& Davis, R.~A. 2002, {Introduction to Time Series and
  Forecasting}, 2nd edn. (New York, NY: Springer)

\bibitem[{{Broderick} {et~al.}(2011){Broderick}, {Fish}, {Doeleman}, \&
  {Loeb}}]{Broderick_2011}
{Broderick}, A.~E., {Fish}, V.~L., {Doeleman}, S.~S., \& {Loeb}, A. 2011, \apj,
  735, 110

\bibitem[{{Broderick} \& {Loeb}(2005)}]{Broderick_Loeb_2005}
{Broderick}, A.~E., \& {Loeb}, A. 2005, \mnras, 363, 353

\bibitem[{{Broderick} \& {Loeb}(2006)}]{Broderick_Loeb_2006}
---. 2006, \mnras, 367, 905

\bibitem[{{Broderick} \& {Loeb}(2009)}]{Broderick_Loeb_2009}
---. 2009, \apjl, 703, L104

\bibitem[{{Broderick} {et~al.}(2009){Broderick}, {Loeb}, \&
  {Narayan}}]{BNL_2009}
{Broderick}, A.~E., {Loeb}, A., \& {Narayan}, R. 2009, \apj, 701, 1357

\bibitem[{{Dexter}(2014)}]{Dexter_2014}
{Dexter}, J. 2014, in IAU Symposium, Vol. 303, IAU Symposium, ed. L.~O.
  {Sjouwerman}, C.~C. {Lang}, \& J.~{Ott}, 298--302

\bibitem[{{Dexter} {et~al.}(2010){Dexter}, {Agol}, {Fragile}, \&
  {McKinney}}]{Dexter_2010}
{Dexter}, J., {Agol}, E., {Fragile}, P.~C., \& {McKinney}, J.~C. 2010, \apj,
  717, 1092

\bibitem[{{Doeleman} {et~al.}(2009{\natexlab{a}}){Doeleman}, {Agol}, {Backer},
  {Baganoff}, {Bower}, {Broderick}, {Fabian}, {Fish}, {Gammie}, {Ho}, {Honman},
  {Krichbaum}, {Loeb}, {Marrone}, {Reid}, {Rogers}, {Shapiro}, {Strittmatter},
  {Tilanus}, {Weintroub}, {Whitney}, {Wright}, \& {Ziurys}}]{Doeleman_2009}
{Doeleman}, S., {et~al.} 2009{\natexlab{a}}, in Astronomy, Vol. 2010,
  astro2010: The Astronomy and Astrophysics Decadal Survey, 68

\bibitem[{{Doeleman} {et~al.}(2009{\natexlab{b}}){Doeleman}, {Fish},
  {Broderick}, {Loeb}, \& {Rogers}}]{Doeleman_Hotspots}
{Doeleman}, S.~S., {Fish}, V.~L., {Broderick}, A.~E., {Loeb}, A., \& {Rogers},
  A.~E.~E. 2009{\natexlab{b}}, \apj, 695, 59

\bibitem[{{Doeleman} {et~al.}(2008){Doeleman}, {Weintroub}, {Rogers},
  {Plambeck}, {Freund}, {Tilanus}, {Friberg}, {Ziurys}, {Moran}, {Corey},
  {Young}, {Smythe}, {Titus}, {Marrone}, {Cappallo}, {Bock}, {Bower},
  {Chamberlin}, {Davis}, {Krichbaum}, {Lamb}, {Maness}, {Niell}, {Roy},
  {Strittmatter}, {Werthimer}, {Whitney}, \& {Woody}}]{Doeleman_2008}
{Doeleman}, S.~S., {et~al.} 2008, \nat, 455, 78

\bibitem[{{Doeleman} {et~al.}(2012){Doeleman}, {Fish}, {Schenck}, {Beaudoin},
  {Blundell}, {Bower}, {Broderick}, {Chamberlin}, {Freund}, {Friberg},
  {Gurwell}, {Ho}, {Honma}, {Inoue}, {Krichbaum}, {Lamb}, {Loeb}, {Lonsdale},
  {Marrone}, {Moran}, {Oyama}, {Plambeck}, {Primiani}, {Rogers}, {Smythe},
  {SooHoo}, {Strittmatter}, {Tilanus}, {Titus}, {Weintroub}, {Wright}, {Young},
  \& {Ziurys}}]{Doeleman_2012}
---. 2012, Science, 338, 355

\bibitem[{{Dolence} {et~al.}(2012){Dolence}, {Gammie}, {Shiokawa}, \&
  {Noble}}]{Dolence_2012}
{Dolence}, J.~C., {Gammie}, C.~F., {Shiokawa}, H., \& {Noble}, S.~C. 2012,
  \apjl, 746, L10

\bibitem[{{Edelson} \& {Krolik}(1988)}]{Edelson_Krolik_1988}
{Edelson}, R.~A., \& {Krolik}, J.~H. 1988, \apj, 333, 646

\bibitem[{{Esin} {et~al.}(1996){Esin}, {Narayan}, {Ostriker}, \&
  {Yi}}]{Esin_1996}
{Esin}, A.~A., {Narayan}, R., {Ostriker}, E., \& {Yi}, I. 1996, \apj, 465, 312

\bibitem[{{Falcke} {et~al.}(2000){Falcke}, {Melia}, \& {Agol}}]{Falcke_2000}
{Falcke}, H., {Melia}, F., \& {Agol}, E. 2000, \apjl, 528, L13

\bibitem[{{Fish} {et~al.}(2009){Fish}, {Doeleman}, {Broderick}, {Loeb}, \&
  {Rogers}}]{Fish_Hotspots}
{Fish}, V.~L., {Doeleman}, S.~S., {Broderick}, A.~E., {Loeb}, A., \& {Rogers},
  A.~E.~E. 2009, \apj, 706, 1353

\bibitem[{{Fish} {et~al.}(2011){Fish}, {Doeleman}, {Beaudoin}, {Blundell},
  {Bolin}, {Bower}, {Chamberlin}, {Freund}, {Friberg}, {Gurwell}, {Honma},
  {Inoue}, {Krichbaum}, {Lamb}, {Marrone}, {Moran}, {Oyama}, {Plambeck},
  {Primiani}, {Rogers}, {Smythe}, {SooHoo}, {Strittmatter}, {Tilanus}, {Titus},
  {Weintroub}, {Wright}, {Woody}, {Young}, \& {Ziurys}}]{Fish_2011}
{Fish}, V.~L., {et~al.} 2011, \apjl, 727, L36

\bibitem[{{Fish} {et~al.}(2014){Fish}, {Johnson}, {Lu}, {Doeleman}, {Bouman},
  {Zoran}, {Freeman}, {Psaltis}, {Narayan}, {Pankratius}, {Broderick}, {Gwinn},
  \& {Vertatschitsch}}]{Fish_2014}
---. 2014, \apj, 795, 134

\bibitem[{{Fishbone} \& {Moncrief}(1976)}]{Fishbone_Moncrief_1976}
{Fishbone}, L.~G., \& {Moncrief}, V. 1976, \apj, 207, 962

\bibitem[{{Gammie} {et~al.}(2003){Gammie}, {McKinney}, \&
  {T{\'o}th}}]{Gammie_2003}
{Gammie}, C.~F., {McKinney}, J.~C., \& {T{\'o}th}, G. 2003, \apj, 589, 444

\bibitem[{{Genzel} {et~al.}(2010){Genzel}, {Eisenhauer}, \&
  {Gillessen}}]{Genzel_2010}
{Genzel}, R., {Eisenhauer}, F., \& {Gillessen}, S. 2010, Reviews of Modern
  Physics, 82, 3121

\bibitem[{{Ghez} {et~al.}(2008){Ghez}, {Salim}, {Weinberg}, {Lu}, {Do}, {Dunn},
  {Matthews}, {Morris}, {Yelda}, {Becklin}, {Kremenek}, {Milosavljevic}, \&
  {Naiman}}]{Ghez_2008}
{Ghez}, A.~M., {et~al.} 2008, \apj, 689, 1044

\bibitem[{{Gillessen} {et~al.}(2009){Gillessen}, {Eisenhauer}, {Trippe},
  {Alexander}, {Genzel}, {Martins}, \& {Ott}}]{Gillessen_2009}
{Gillessen}, S., {Eisenhauer}, F., {Trippe}, S., {Alexander}, T., {Genzel}, R.,
  {Martins}, F., \& {Ott}, T. 2009, \apj, 692, 1075

\bibitem[{{Gillessen} {et~al.}(2012){Gillessen}, {Genzel}, {Fritz}, {Quataert},
  {Alig}, {Burkert}, {Cuadra}, {Eisenhauer}, {Pfuhl}, {Dodds-Eden}, {Gammie},
  \& {Ott}}]{Gillessen_2012}
{Gillessen}, S., {et~al.} 2012, \nat, 481, 51

\bibitem[{{Goodman} \& {Narayan}(1989)}]{GoodmanNarayan89}
{Goodman}, J., \& {Narayan}, R. 1989, \mnras, 238, 995

\bibitem[{{Gwinn} {et~al.}(2014){Gwinn}, {Kovalev}, {Johnson}, \&
  {Soglasnov}}]{Gwinn_2014}
{Gwinn}, C.~R., {Kovalev}, Y.~Y., {Johnson}, M.~D., \& {Soglasnov}, V.~A. 2014,
  \apjl, 794, L14

\bibitem[{{Hamaus} {et~al.}(2009){Hamaus}, {Paumard}, {M{\"u}ller},
  {Gillessen}, {Eisenhauer}, {Trippe}, \& {Genzel}}]{Hamaus_2009}
{Hamaus}, N., {Paumard}, T., {M{\"u}ller}, T., {Gillessen}, S., {Eisenhauer},
  F., {Trippe}, S., \& {Genzel}, R. 2009, \apj, 692, 902

\bibitem[{{Johnson} {et~al.}(2014){Johnson}, {Fish}, {Doeleman}, {Broderick},
  {Wardle}, \& {Marrone}}]{Johnson_2014}
{Johnson}, M.~D., {Fish}, V.~L., {Doeleman}, S.~S., {Broderick}, A.~E.,
  {Wardle}, J.~F.~C., \& {Marrone}, D.~P. 2014, \apj, 794, 150

\bibitem[{{Johnson} \& {Gwinn}(2015)}]{Johnson_Gwinn_2015}
{Johnson}, M.~D., \& {Gwinn}, C.~R. 2015, ArXiv e-prints

\bibitem[{{Kardashev} {et~al.}(2013){Kardashev}, {Khartov}, {Abramov},
  {Avdeev}, {Alakoz}, {Aleksandrov}, {Ananthakrishnan}, {Andreyanov},
  {Andrianov}, {Antonov}, {Artyukhov}, {Arkhipov}, {Baan}, {Babakin},
  {Babyshkin}, {Bartel'}, {Belousov}, {Belyaev}, {Berulis}, {Burke},
  {Biryukov}, {Bubnov}, {Burgin}, {Busca}, {Bykadorov}, {Bychkova},
  {Vasil'kov}, {Wellington}, {Vinogradov}, {Wietfeldt}, {Voitsik},
  {Gvamichava}, {Girin}, {Gurvits}, {Dagkesamanskii}, {D'Addario},
  {Giovannini}, {Jauncey}, {Dewdney}, {D'yakov}, {Zharov}, {Zhuravlev},
  {Zaslavskii}, {Zakhvatkin}, {Zinov'ev}, {Ilinen}, {Ipatov}, {Kanevskii},
  {Knorin}, {Casse}, {Kellermann}, {Kovalev}, {Kovalev}, {Kovalenko}, {Kogan},
  {Komaev}, {Konovalenko}, {Kopelyanskii}, {Korneev}, {Kostenko}, {Kotik},
  {Kreisman}, {Kukushkin}, {Kulishenko}, {Cooper}, {Kut'kin}, {Cannon},
  {Larionov}, {Lisakov}, {Litvinenko}, {Likhachev}, {Likhacheva}, {Lobanov},
  {Logvinenko}, {Langston}, {McCracken}, {Medvedev}, {Melekhin}, {Menderov},
  {Murphy}, {Mizyakina}, {Mozgovoi}, {Nikolaev}, {Novikov}, {Novikov},
  {Oreshko}, {Pavlenko}, {Pashchenko}, {Ponomarev}, {Popov}, {Pravin-Kumar},
  {Preston}, {Pyshnov}, {Rakhimov}, {Rozhkov}, {Romney}, {Rocha}, {Rudakov},
  {R{\"a}is{\"a}nen}, {Sazankov}, {Sakharov}, {Semenov}, {Serebrennikov},
  {Schilizzi}, {Skulachev}, {Slysh}, {Smirnov}, {Smith}, {Soglasnov},
  {Sokolovskii}, {Sondaar}, {Stepan'yants}, {Turygin}, {Turygin}, {Tuchin},
  {Urpo}, {Fedorchuk}, {Finkel'shtein}, {Fomalont}, {Fejes}, {Fomina},
  {Khapin}, {Tsarevskii}, {Zensus}, {Chuprikov}, {Shatskaya}, {Shapirovskaya},
  {Sheikhet}, {Shirshakov}, {Schmidt}, {Shnyreva}, {Shpilevskii}, {Ekers}, \&
  {Yakimov}}]{Kardashev_2013}
{Kardashev}, N.~S., {et~al.} 2013, Astronomy Reports, 57, 153

\bibitem[{{Lo} {et~al.}(1998){Lo}, {Shen}, {Zhao}, \& {Ho}}]{Lo_1998}
{Lo}, K.~Y., {Shen}, Z.-Q., {Zhao}, J.-H., \& {Ho}, P.~T.~P. 1998, \apjl, 508,
  L61

\bibitem[{{Loeb} \& {Waxman}(2007)}]{Loeb_Waxman_2007}
{Loeb}, A., \& {Waxman}, E. 2007, \jcap, 3, 11

\bibitem[{{Lu} {et~al.}(2014){Lu}, {Broderick}, {Baron}, {Monnier}, {Fish},
  {Doeleman}, \& {Pankratius}}]{Lu_2014}
{Lu}, R.-S., {Broderick}, A.~E., {Baron}, F., {Monnier}, J.~D., {Fish}, V.~L.,
  {Doeleman}, S.~S., \& {Pankratius}, V. 2014, \apj, 788, 120

\bibitem[{{Lu} {et~al.}(2011){Lu}, {Krichbaum}, {Eckart}, {K{\"o}nig},
  {Kunneriath}, {Witzel}, {Witzel}, \& {Zensus}}]{Lu_2011}
{Lu}, R.-S., {Krichbaum}, T.~P., {Eckart}, A., {K{\"o}nig}, S., {Kunneriath},
  D., {Witzel}, G., {Witzel}, A., \& {Zensus}, J.~A. 2011, \aap, 525, A76

\bibitem[{{Marrone} {et~al.}(2007){Marrone}, {Moran}, {Zhao}, \&
  {Rao}}]{Marrone_2007}
{Marrone}, D.~P., {Moran}, J.~M., {Zhao}, J.-H., \& {Rao}, R. 2007, \apjl, 654,
  L57

\bibitem[{{McCourt} \& {Madigan}(2015)}]{McCourt_Madigan_2015}
{McCourt}, M., \& {Madigan}, A.-M. 2015, ArXiv e-prints

\bibitem[{{Mo{\'s}cibrodzka} {et~al.}(2009){Mo{\'s}cibrodzka}, {Gammie},
  {Dolence}, {Shiokawa}, \& {Leung}}]{Moscibrodzka_2009}
{Mo{\'s}cibrodzka}, M., {Gammie}, C.~F., {Dolence}, J.~C., {Shiokawa}, H., \&
  {Leung}, P.~K. 2009, \apj, 706, 497

\bibitem[{{Narayan} \& {Goodman}(1989)}]{NarayanGoodman89}
{Narayan}, R., \& {Goodman}, J. 1989, \mnras, 238, 963

\bibitem[{{Narayan} \& {Yi}(1994)}]{Narayan_Yi_1994}
{Narayan}, R., \& {Yi}, I. 1994, \apjl, 428, L13

\bibitem[{{Noble} {et~al.}(2006){Noble}, {Gammie}, {McKinney}, \& {Del
  Zanna}}]{Noble_2006}
{Noble}, S.~C., {Gammie}, C.~F., {McKinney}, J.~C., \& {Del Zanna}, L. 2006,
  \apj, 641, 626

\bibitem[{{Noble} {et~al.}(2007){Noble}, {Leung}, {Gammie}, \&
  {Book}}]{Noble_2007}
{Noble}, S.~C., {Leung}, P.~K., {Gammie}, C.~F., \& {Book}, L.~G. 2007,
  Classical and Quantum Gravity, 24, 259

\bibitem[{{Pfuhl} {et~al.}(2015){Pfuhl}, {Gillessen}, {Eisenhauer}, {Genzel},
  {Plewa}, {Ott}, {Ballone}, {Schartmann}, {Burkert}, {Fritz}, {Sari},
  {Steinberg}, \& {Madigan}}]{Pfuhl_2015}
{Pfuhl}, O., {et~al.} 2015, \apj, 798, 111

\bibitem[{{Psaltis} {et~al.}(2015){Psaltis}, {Narayan}, {Fish}, {Broderick},
  {Loeb}, \& {Doeleman}}]{Psaltis_2015}
{Psaltis}, D., {Narayan}, R., {Fish}, V.~L., {Broderick}, A.~E., {Loeb}, A., \&
  {Doeleman}, S.~S. 2015, \apj, 798, 15

\bibitem[{{Quataert} \& {Gruzinov}(2000)}]{Quataert_Gruzinov_2000}
{Quataert}, E., \& {Gruzinov}, A. 2000, \apj, 545, 842

\bibitem[{{Roberts} {et~al.}(1994){Roberts}, {Wardle}, \& {Brown}}]{RWB94}
{Roberts}, D.~H., {Wardle}, J.~F.~C., \& {Brown}, L.~F. 1994, \apj, 427, 718

\bibitem[{{Sault} {et~al.}(1997){Sault}, {Oosterloo}, {Dulk}, \&
  {Leblanc}}]{Sault_1997}
{Sault}, R.~J., {Oosterloo}, T., {Dulk}, G.~A., \& {Leblanc}, Y. 1997, \aap,
  324, 1190

\bibitem[{{Shcherbakov} \& {McKinney}(2013)}]{Shcherbakov_2013}
{Shcherbakov}, R.~V., \& {McKinney}, J.~C. 2013, \apjl, 774, L22

\bibitem[{{Shcherbakov} {et~al.}(2012){Shcherbakov}, {Penna}, \&
  {McKinney}}]{Shcherbakov_2012}
{Shcherbakov}, R.~V., {Penna}, R.~F., \& {McKinney}, J.~C. 2012, \apj, 755, 133

\bibitem[{{Shen} {et~al.}(2005){Shen}, {Lo}, {Liang}, {Ho}, \&
  {Zhao}}]{Shen_2005}
{Shen}, Z.-Q., {Lo}, K.~Y., {Liang}, M.-C., {Ho}, P.~T.~P., \& {Zhao}, J.-H.
  2005, \nat, 438, 62

\bibitem[{{Shiokawa}(2013)}]{Shiokawa_Thesis}
{Shiokawa}, H. 2013, PhD thesis, University of Illinois at Urbana-Champaign

\bibitem[{{Takahashi}(2004)}]{Takahashi_2004}
{Takahashi}, R. 2004, \apj, 611, 996

\bibitem[{{Thompson} {et~al.}(2001){Thompson}, {Moran}, \& {Swenson}}]{TMS}
{Thompson}, A.~R., {Moran}, J.~M., \& {Swenson}, Jr., G.~W. 2001,
  {Interferometry and Synthesis in Radio Astronomy, 2nd Edition}

\bibitem[{{Vincent} {et~al.}(2011){Vincent}, {Paumard}, {Perrin}, {Mugnier},
  {Eisenhauer}, \& {Gillessen}}]{Vincent_2011}
{Vincent}, F.~H., {Paumard}, T., {Perrin}, G., {Mugnier}, L., {Eisenhauer}, F.,
  \& {Gillessen}, S. 2011, \mnras, 412, 2653

\bibitem[{{Wild} {et~al.}(2009){Wild}, {Kardashev}, {Likhachev}, {Babakin},
  {Arkhipov}, {Vinogradov}, {Andreyanov}, {Fedorchuk}, {Myshonkova},
  {Alexsandrov}, {Novokov}, {Goltsman}, {Cherepaschuk}, {Shustov}, {Vystavkin},
  {Koshelets}, {Vdovin}, {de Graauw}, {Helmich}, {Vd Tak}, {Shipman},
  {Baryshev}, {Gao}, {Khosropanah}, {Roelfsema}, {Barthel}, {Spaans}, {Mendez},
  {Klapwijk}, {Israel}, {Hogerheijde}, {Vd Werf}, {Cernicharo},
  {Martin-Pintado}, {Planesas}, {Gallego}, {Beaudin}, {Krieg}, {Gerin},
  {Pagani}, {Saraceno}, {di Giorgio}, {Cerulli}, {Orfei}, {Spinoglio},
  {Piazzo}, {Liseau}, {Belitsky}, {Cherednichenko}, {Poglitsch}, {Raab},
  {Guesten}, {Klein}, {Stutzki}, {Honingh}, {Benz}, {Murphy}, {Trappe}, \&
  {R{\"a}is{\"a}nen}}]{Wild_2009}
{Wild}, W., {et~al.} 2009, Experimental Astronomy, 23, 221

\bibitem[{{Yuan} \& {Narayan}(2014)}]{Yuan_2014}
{Yuan}, F., \& {Narayan}, R. 2014, \araa, 52, 529

\bibitem[{{Yuan} {et~al.}(2003){Yuan}, {Quataert}, \& {Narayan}}]{Yuan_2003}
{Yuan}, F., {Quataert}, E., \& {Narayan}, R. 2003, \apj, 598, 301

\end{thebibliography}

\end{document}